\documentclass[aps,superscriptaddress,prc,twocolumn,nofootinbib]{revtex4}

\usepackage{amsmath,amssymb,amsfonts}
\usepackage{graphicx}% Include figure files
\usepackage{dcolumn}% Align table columns on decimal point
\usepackage{bm}% bold math
\usepackage{hyperref}% add hypertext capabilities

\begin{document}

\title{Jet shape and redistribution of the lost energy from jets in Pb+Pb collisions at the LHC in a multiphase transport model}%

\author{Ao Luo}
%\email{aluo@mails.ccnu.edu.cn}
\affiliation{Key Laboratory of Quark and Lepton Physics (MOE) and Institute of Particle Physics,
	\\ Central China Normal University, Wuhan 430079, China}

\author{Ya-Xian Mao}
%\email{yaxian.mao@ccnu.edu.cn}
\affiliation{Key Laboratory of Quark and Lepton Physics (MOE) and Institute of Particle Physics,
	\\ Central China Normal University, Wuhan 430079, China}

\author{Guang-You Qin}
%\email{guangyou.qin@mail.ccnu.edu.cn}
\affiliation{Key Laboratory of Quark and Lepton Physics (MOE) and Institute of Particle Physics,
	\\ Central China Normal University, Wuhan 430079, China}

\author{En-Ke Wang}
%\email{wangek@scnu.edu.cn}
\affiliation{Guangdong Provincial Key Laboratory of Nuclear Science, Institute of Quantum Matter,
    \\ South China Normal University, Guangzhou 510006, China}
\affiliation{Key Laboratory of Quark and Lepton Physics (MOE) and Institute of Particle Physics,
	\\ Central China Normal University, Wuhan 430079, China}

\author{Han-Zhong Zhang}
%\email{zhanghz@mail.ccnu.edu.cn}
\affiliation{Key Laboratory of Quark and Lepton Physics (MOE) and Institute of Particle Physics,
	\\ Central China Normal University, Wuhan 430079, China}

\begin{abstract}

Jet-medium interaction involves two important effects: jet energy loss and medium response.
The search for jet-induced medium excitations is one of the hot topics in jet quenching study in relativistic nuclear collisions.
In this work, we perform a systematic study on how the lost energy from hard jets evolves with the bulk medium and redistributes in the final state of heavy-ion collisions via a multi-phase transport model.
In particular, the ($\Delta \eta, \Delta \phi$) distribution of charged particles with respect to the jet axis and jet shape function are studied for various Pb+Pb collision centralities and for different transverse momentum intervals of charged particles.
Our numerical result shows a strong enhancement of soft particles at large angles for Pb+Pb collisions relative to p+p collisions at the LHC, qualitatively consistent with recent CMS data.
This indicates that a significant fraction of the lost energy from hard jets is carried by soft particles at large angles away from the jet axis.

\end{abstract}

\maketitle

%\tableofcontents

\section{Introduction \label{sec:introduction}}

Ultra-relativistic heavy-ion collisions can create extremely hot and dense nuclear matter, commonly referred to as Quark-Gluon Plasma (QGP) \cite{Gyulassy:2004zy, Adcox:2004mh, Arsene:2004fa, Back:2004je, Adams:2005dq, Muller:2006ee, Muller:2012zq}.
High transverse momentum ($p_\mathrm {T}$) jet partons produced in early-stage hard scatterings interact with the hot QCD matter via medium-induced gluon bremsstrahlung and elastic collisions before hadronizing into hadrons.
The interaction between jets and QGP medium usually leads to parton energy loss, transverse momentum broadening, etc.
These phenomena are collectively referred to as jet quenching \cite{Wang:1991xy, Qin:2015srf, Cao:2020wlm}.
Jet quenching provides a powerful tool to probe the novel properties of QGP \cite{Bernhard:2019bmu, Everett:2020xug, Burke:2013yra, Cao:2021keo}.
One important evidence of jet quenching is the suppression of large transverse momentum hadrons produced in heavy-ion collisions \cite{Adcox:2001jp, Adler:2002xw, Aamodt:2010jd}, which has been observed at Relativistic Heavy Ion Collider (RHIC) and the Large Hadron Collider (LHC).
The central goal of jet quenching study is to understand in detail the interaction mechanisms between jets and QCD medium, to study how different effects manifest in the final states of the collisions \cite{Qin:2007rn, Zhang:2007ja, Zhang:2009rn, Bass:2008rv, Qin:2009bk, Chen:2011vt, Chen:2016vem, Chen:2016cof, Zhang:2018urd}, and to probe the structure and transport properties of QGP at various scales.

Another important aspect of jet-medium interaction is the medium response to jet propagation \cite{CasalderreySolana:2004qm, Chaudhuri:2005vc, Ruppert:2005uz, Qin:2009uh, Neufeld:2009ep}.
The lost energy and momentum from hard jets may induce medium excitations, which may affect the space-time evolution of the bulk matter and many jet-related observables \cite{Andrade:2014swa, Schulc:2014jma, Tachibana:2017syd, Chang:2019sae, Chen:2017zte, Chen:2020tbl, Yang:2021iib, Luo:2018pto, Park:2018acg, KunnawalkamElayavalli:2017hxo, Brewer:2017fqy}.
For example, fast jet partons may induce Mach cone like excitation in the medium \cite{CasalderreySolana:2004qm, Ruppert:2005uz}.
If observed, it can provide a direct probe to the speed of sound of the hot QCD medium.
The detailed structure of Mach cone is also sensitive to many transport properties of the medium such as the specific shear viscosity \cite{Neufeld:2008dx, Bouras:2014rea}.
In the dynamically evolving media, however, the detection of Mach-cone signal is quite difficult since the collective flow of the expanding media can distort the Mach cone structure induced by hard jets \cite{Renk:2005si, Ma:2010dv, Bouras:2014rea}.
The search for jet-induced medium response effect is still an on-going hot topic in jet quenching study in heavy-ion collisions \cite{Tachibana:2017syd, Chang:2019sae, Chen:2017zte, Chen:2020tbl, Yang:2021iib}.

Fully reconstructed jets and their nuclear modification provide many new insights into the study of jet energy loss and medium response in relativistic heavy-ion collisions.
As for full jets, one has to consider not only the interaction between the leading partons and the medium, but also the evolution of the shower partons of the jets \cite{Qin:2010mn}.
When jets propagate through the QGP, jet-medium interaction not only reduces the total energy of the reconstructed jets, but also change the energy and momentum distributions among the jet constituents \cite{Chang:2016gjp, Tachibana:2017syd, Chang:2019sae}.
There have been many theoretical and experimental studies on the suppression of jet yield and the nuclear modification of jet structure \cite{Chatrchyan:2013kwa, Chatrchyan:2014ava, Aad:2014wha, Khachatryan:2015lha, Young:2011qx, Dai:2012am, Wang:2013cia, Blaizot:2013hx, Mehtar-Tani:2014yea, Cao:2017qpx, Kang:2017frl, He:2018xjv, Chang:2016gjp, Casalderrey-Solana:2016jvj, KunnawalkamElayavalli:2017hxo, Brewer:2017fqy, Chien:2016led, Milhano:2017nzm, Chang:2019sae}.
On the other hand, it is very interesting to use full jets to investigate how the lost energy from the jets evolves in the QGP medium and redistributes in the final states of nuclear collisions.
Reference \cite{Tachibana:2017syd} has shown that jet-deposited energy can be transported to very large angles away from the jet direction via  hydrodynamic medium evolution.
In Ref. \cite{Gao:2016ldo}, we have studied the overall momentum balance and the redistribution of lost energy for dijet events \cite{CMS:2011iwn} within the framework of AMPT model and found that elastic collisions between jets and medium partons can play very important roles in transporting the lost energy from jets to very large angles.

To further investigate where the lost energy redistributes in the final state, one may study the detailed distributions of final-state particles with different momenta around the reconstructed jets up to very large angles away from the jet axis.
Recently, CMS Collaboration has measured the correlations of charged particles with respect to jet axis as a function of relative pseudorapidty ($\Delta \eta$), relative azimuthal angle ($\Delta \phi$), and relative radial distance $\Delta r = \sqrt{(\Delta\eta)^2 + (\Delta\phi)^2}$ for Pb+Pb collisions and p+p collision at $\sqrt {s_ \mathrm{NN}}= 5.02$~TeV at the LHC \cite{CMS:2018zze}.
The ($\Delta \eta, \Delta \phi$) distribution of charged particle yield and jet shape $\rho(\Delta r)$ have been measured for different Pb+Pb collision centralities and for different transverse momentum intervals of charged particles.
Such differential measurements provide a unique opportunity for studying the redistribution of the lost energy from hard jets due to jet-medium interaction.

In this work, we use a multi-phase transport (AMPT) model \cite{Lin:2004en} to study jet quenching and the redistribution of jet energy loss in high-energy nuclear collisions.
We follow the CMS Collaboration and perform a detailed investigation on the correlations of charged particles with respect to jet axis as a function of $\Delta \eta$ and $\Delta \phi$ and jet shape function $\rho(\Delta r)$ up to large radial distance ($\Delta r \sim 1$) for p+p and Pb+Pb collisions at $\sqrt {s_\mathrm{NN}} = 5.02$ TeV.
Our work constitutes an important study on understanding the nuclear modification of jet shower and the medium response to jets in relativistic heavy-ion collisions.

The paper is organized as follows.
In Sec \ref{sec:framework}, we provide a brief introduction to the AMPT model.
Then we describe the observables and the detailed analysis procedures in jet reconstruction and jet-particle correlations.
Numerical results are presented in Sec \ref{sec:results}.
Sec \ref{sec:summary} contains the summary of this work.

\section{The AMPT model and observables \label{sec:framework}}

\subsection{The AMPT model}

In this work, we use the AMPT model with string melting mechanism \cite{Zhang:2005ni, Ma:2011uma, Ma:2010dv} to simulate jet-medium interaction in high-energy heavy-ion collisions.
Generally, the AMPT model consists of four main stages: initial condition, parton cascade, hadronization, and hadronic rescatterings.

\begin{itemize}

	\item The initial condition of the AMPT model is provided by the HIJING model \cite{Wang:1991hta, Gyulassy:1994ew}, which generates the initial spatial and momentum distributions of the matter participants, including minijet partons and soft string excitations. In order to study jet quenching, a jet trigger technique in the HIJING model is applied in the AMPT model to increase the efficiency. Several hard dijet production channels are included: $qq \rightarrow qq$, $q\bar{q}\rightarrow q\bar{q}$, $q\bar{q}\rightarrow gg$, $qg\rightarrow qg$, $gg\rightarrow q\bar{q}$, and $gg\rightarrow gg$ \cite{Sjostrand:1993yb}.

	\item In the AMPT model, the dynamical evolution of partons is simulated via Zhang's parton cascade (ZPC) model \cite{Zhang:1997ej}. ZPC model describes elastic collisions among the medium partons and jet partons.
The partonic interaction cross section $\mathrm{\sigma}$ is determined by the value of strong coupling constant and the Debye screening mass. Only elastic collisions are included in the AMPT model at present.

	\item Hadronization process starts when the partonic system freezes out. A quark coalescence model is used to combine partons into hadrons \cite{Lin:2001zk}.

	\item The dynamics of hadronic interactions is simulated via a relativistic transport (ART) model \cite{Li:1995pra}, which includes baryon-baryon, baryon-meson, meson-meson elastic and inelastic scatterings.
\end{itemize}

The AMPT model has successfully described many jet observables, such as $\gamma$-jet imbalance \cite{Ma:2013bia}, dijet asymmetry \cite{Ma:2013pha, Gao:2016ldo}, jet fragmentation function \cite{Ma:2013gga,Ma:2013yoa} and jet shape \cite{Ma:2013uqa}. Here we use the correlations of charged particles and reconstructed jets to study the distribution of charged particles and jet shape up to very large angles with respect to jet axis.

\subsection{Jet observables and the analysis method}

The interaction between high-$p_\mathrm{T}$ partons and QGP medium leads to energy loss from jet cone which causes the reduction of the jet yield at a given transverse momentum ($p_\mathrm{T}$).
The suppression of the inclusive jet spectra in $A+A$ collisions relative to p+p collisions can be quantified by the nuclear modification factor $R_\mathrm{AA} (\mathrm {p^{jet}_T})$, which is is defined as the ratio of jet $\mathrm {p^{jet}_T}$ spectra in $A+A$ collisions to that in p+p collisions scaled with the number of binary nucleon-nucleon collisions $N_\mathrm{coll}$:
\begin{equation}
R_\mathrm{AA} (\mathrm {p^{jet}_T}) =  \frac{1}{\mathrm N_\mathrm{coll}}\frac{  { \mathrm{d^2 N^{jet}_{AA}}}/{ \mathrm {dp^{jet}_T dy} } }{ { \mathrm{d^2 N^{jet}_{pp}}}/{ \mathrm {dp^{jet}_T dy} }}.
\end{equation}
We apply the FASTJET framework \cite{Cacciari:2011ma} with the anti-$k_\mathrm{T}$ algorithm to reconstruct jets in Pb+Pb and p+p events.
Note that our simulation of p+p events with AMPT also contains four steps listed in Sec. IIA except that the collision system is much smaller.
The jet cone size is taken to be $R=0.4$.
The contribution from the background is subtracted using a variant of the iterative ``noise/pedestal subtraction'' technique as described in Ref. \cite{Kodolova:2007hd}.
In Section \ref{sec:results}, we will show the numerical results for $R_\mathrm{AA} (\mathrm {p^{jet}_T})$ and compare to the ATLAS data \cite{ATLAS:2018gwx}.

Jet shape $\rho (\Delta r)$ measures the transverse momentum distribution of the jet along the radial distance $\Delta r$ with respect to the jet axis.
The nuclear modification of jet shape is sensitive to various jet-medium interaction mechanisms as well as jet-induced medium excitations.
To define the jet shapes $\rho (\Delta r)$, one first evaluates the distribution of charged particles weighted by their $p_\mathrm T$ values as a function of $\Delta r$ and define the transverse momentum profile of the jet $ P(\Delta r)$ as follows:
\begin{equation}
	\label{equation:P}
	{P}(\Delta r)=\frac{1}{\delta \mathrm{r}}\frac{1}{N_{\mathrm{jets}}} \Sigma_{\mathrm{jets}}\Sigma_{\mathrm{particles} \in (\Delta \mathrm{r_a},\Delta \mathrm{r_b})}p_{\mathrm{T}},
\end{equation}
where the bin size $\delta r = \Delta r_b - \Delta r_a$.
Then one normalizes the above jet transverse momentum profile to unity within $\Delta r \in [0, 1]$ to obtain the jet shape function $\rho (\Delta r)$:
\begin{equation}
	\label{equation:rho}
	\rho (\Delta \mathrm{r})=\frac{{P}(\Delta r)}
	{\frac{1}{N_{\mathrm{jets}}}\Sigma_{\mathrm{jets}}\Sigma_{\mathrm{particles}}p_{\mathrm{T}}}.
\end{equation}

The particle and energy-momentum distributions around the jets can be studied via the correlations between charged particles and reconstructed jets.
For this purpose, CMS Collaboration has constructed a two-dimensional charged particle number density $\frac{\mathrm d^2N}{\mathrm d\Delta\phi \mathrm d\Delta\eta}$ per trigger jet, where $\mathrm{\Delta \eta}$ and $\mathrm{\Delta \phi}$ are relative pseudo-rapidity and azimuthal angle with respect to the jet axis.
The particle number density can be studied for different centralities and for different transverse momenta of charged particles.
In this work, the above jet-particle correlations in Pb+Pb collisions are studied in four centrality intervals: $0-10\%$ (most central), $10-30\%$, $30-50\%$, and $50-100\%$ (most peripheral).
We also divide the charged particles into several transverse momentum bins, bounded by 0.7, 1, 2, 3, 4, 6, 8, 12, 16, 20 and 300 GeV.
In this sense, we will construct the three-dimensional distribution $\frac{\mathrm dN}{\mathrm dp_{\rm T} \mathrm d\Delta \eta \mathrm d\Delta \phi}$ of charged particles around the jets for different centralities, which provides a very detailed description on the particle and energy-momentum distributions around the fully-reconstructed jets produced in high-energy nuclear collisions.

\begin{figure}
	\centering
	\includegraphics[width=1\linewidth]{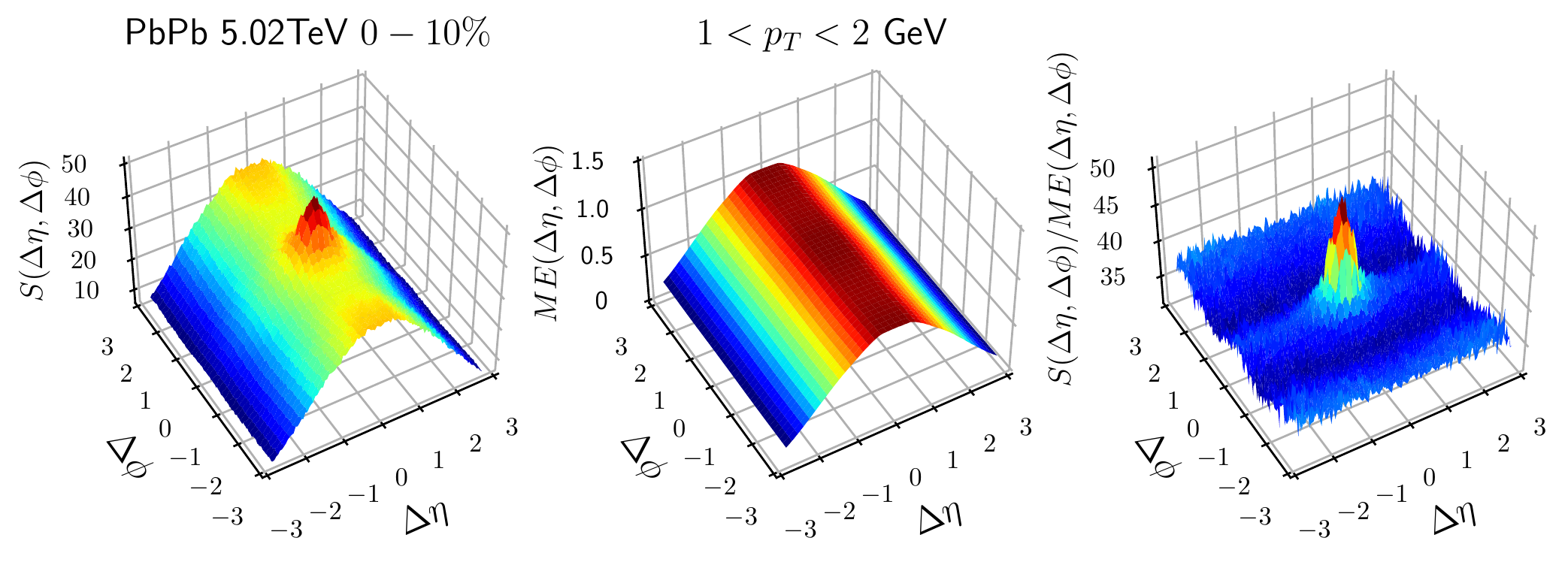}
	\caption{Jet-particle correlations using charged particles with $1 < p\mathrm{_T} < 2$ GeV in central $0-10\%$ Pb+Pb collisions. The left panel is for signal pair distribution $S(\mathrm{\Delta \eta},\mathrm{\Delta \phi})$, the middle for normalized mixed-event pair distribution $ME(\mathrm{\Delta \eta},\mathrm{\Delta \phi})$, and the right for acceptance-corrected per-trigger-jet associated particle yield.}
	\label{fig:Mixed_Event}
\end{figure}

Following the procedure of CMS Collaboration, we construct jet-particle correlations for jets with $|\eta| < 1.6$ and charged particles with $|\eta| < 2.4$.
To correct the limited acceptance, we use a mixed-event method \cite{Khachatryan:2016erx, Khachatryan:2016tfj}.
A mixed-event distribution is constructed by creating jet-particle pairs using jets from the jet-triggered event and particles from another event.
After the acceptance correction, the two-dimensional particle yield (distribution) per trigger jet is obtained as follows:
\begin{equation}
	\frac{1}{N_\mathrm{jets}} \frac{\mathrm{d}^2N}{\mathrm{d}\mathrm{\Delta\eta} \mathrm{d}\mathrm{\Delta \phi}}
	= \frac{ME(0,0)}{ME(\mathrm{\Delta \eta},\mathrm{\Delta \phi})} S(\mathrm{\Delta \eta},\mathrm{\Delta \phi}),
\label{eq:2D_dist}
\end{equation}
where $N_\mathrm{jets}$ denotes the total number of jets satisfying the selection criteria.
The signal pair distribution $S(\mathrm{\Delta \eta},\mathrm{\Delta \phi})$ represents the yield of jet-particles pairs from the same jet-triggered event, normalized by $N_\mathrm{jets}$,
\begin{equation}
	S(\mathrm{\Delta \eta},\mathrm{\Delta \phi}) =\frac{1}{N_\mathrm{jets}} \frac{\mathrm{d}^2N^\mathrm{same}}{\mathrm{d}\mathrm{\Delta\eta} \mathrm{d}\mathrm{\Delta \phi}}.
\end{equation}
The mixed-event pair distribution $ME(\mathrm{\Delta \eta},\mathrm{\Delta \phi})$ is
\begin{equation}
	ME(\mathrm{\Delta \eta},\mathrm{\Delta \phi}) =
	\frac{1}{N_\mathrm{jets}} \frac{\mathrm{d}^2N^\mathrm{mix}}{\mathrm{d}\mathrm{\Delta\eta} \mathrm{d}\mathrm{\Delta \phi}}.
\end{equation}
The ratio ${ME(0,0)}/{ME(\mathrm{\Delta \eta},\mathrm{\Delta \phi})}$ is the normalized correction factor.

As an example, the above procedure of constructing jet-particle correlations is illustrated in Fig. \ref{fig:Mixed_Event} using $0-10\%$ Pb+Pb collisions with charged particles in $1<p_\mathrm{T} <2$ GeV.
First, we apply the FASTJET algorithms and the noise/pedestal subtraction technique to find the jet location and the associated charged particles around the jet to construct the signal pair distribution $S(\Delta \eta, \Delta \phi)$, as shown in the left panel of Fig. \ref{fig:Mixed_Event}.
We use the mixed-event method to construct the mixed-event pair distribution $ME(\Delta \eta, \Delta \phi)$, as shown in the middle panel.
Then, we use Eq. \ref{eq:2D_dist} to obtain the acceptance-corrected two-dimensional particle yield per trigger jet, as shown in right panel of Fig. \ref{fig:Mixed_Event}.
After the acceptance correction, we use the side-band method to subtract the background from uncorrelated pairs and long-range correlations.
Following the CMS Collaboration, this background is estimated using $\Delta \phi$ distribution averaged over $1.5 < |\Delta \eta| < 2.5$, and then subtracted to get the final corrected jet-particle correlations.

\section{Numerical Results \label{sec:results}}

\begin{figure}
	\centering
	\includegraphics[width=1\linewidth]{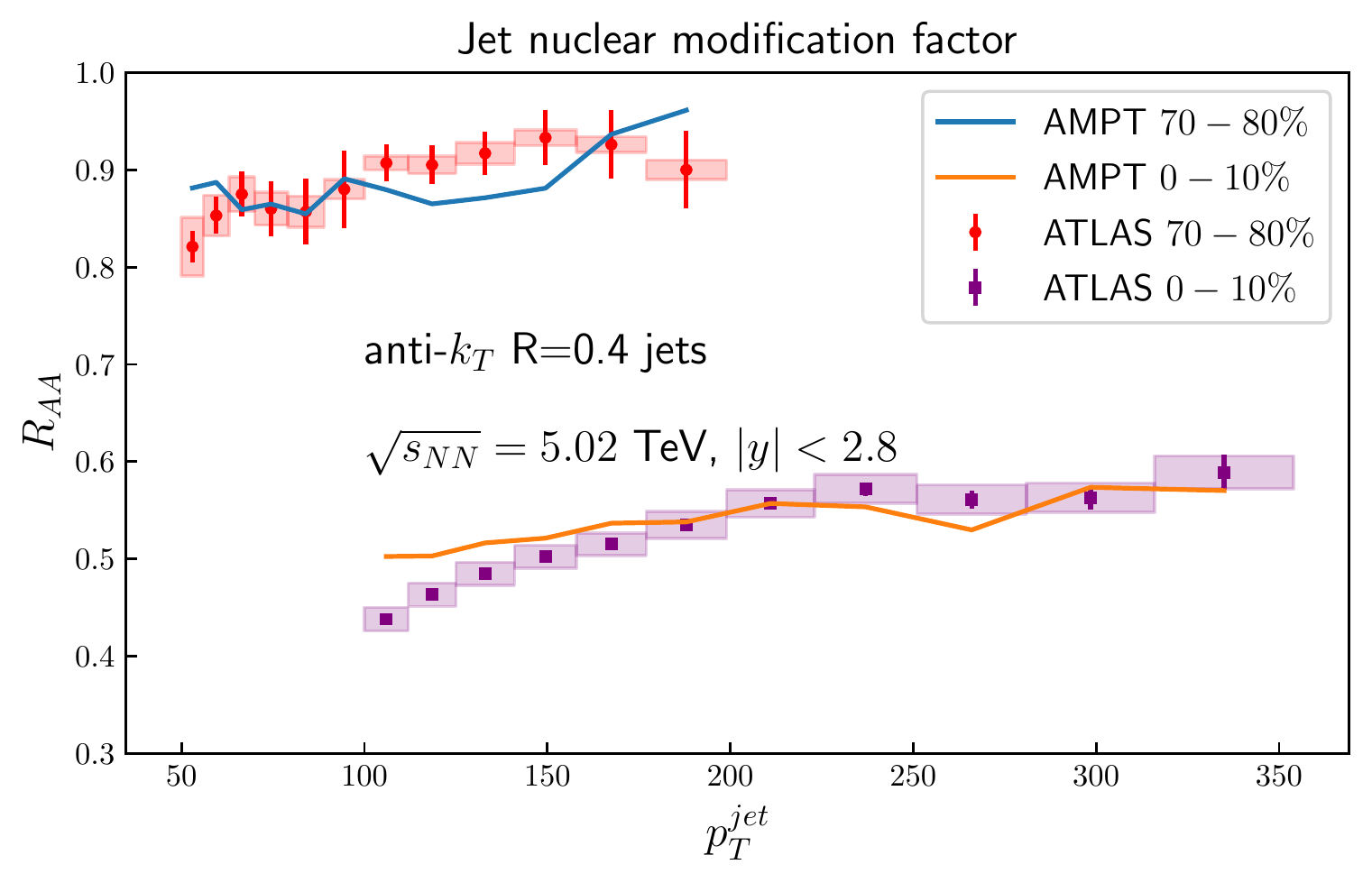}
	\caption{Nuclear modification factor $R_\mathrm{AA}(\mathrm {p^{jet}_T})$ as a function of $p_\mathrm{T}^\mathrm{jet}$ from the AMPT simulation compared with the ATLAS data \cite{ATLAS:2018gwx} for jets with $R = 0.4$ and $\left| y \right| < 2.8$ in central and peripheral Pb+Pb collisions at $\sqrt s = 5.02$ GeV.}
	\label{fig:RAA}
\end{figure}

\begin{figure*}
	\centering
	\includegraphics[width=0.72\linewidth]{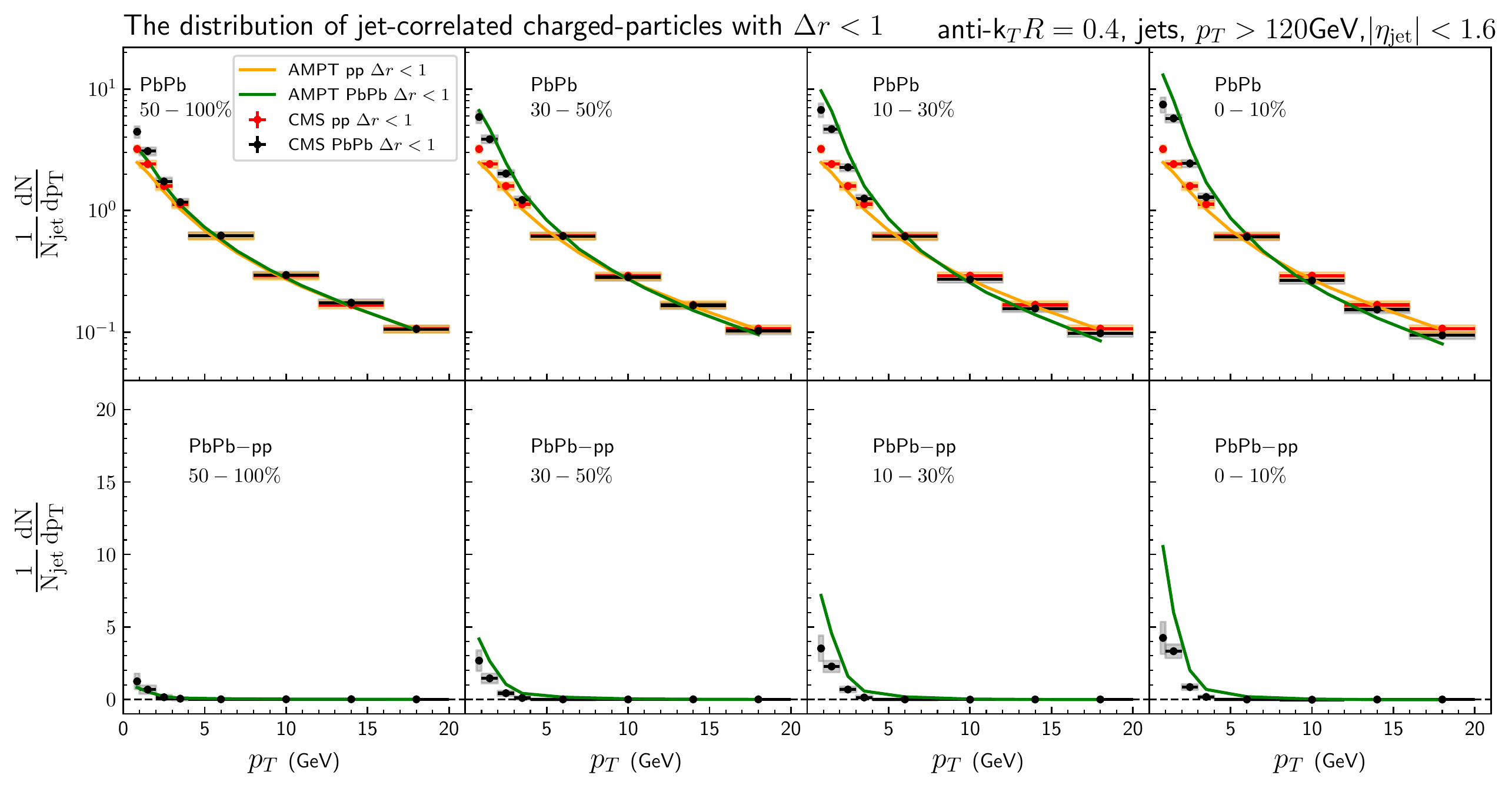}
	\caption{Jet-correlated charged particle yield distribution around the jets in the region $\mathrm{\Delta r<1}$ as a function of charge particle $p_T$ for Pb+Pb and p+p collisions from the AMPT model and compare to CMS data \cite{CMS:2018zze} (upper panels). The differences between Pb+Pb and p+p are shown in the lower panels.
Four Pb+Pb collision centralities are shown: $0-10\%, 10-30\%, 30-50\%$, and $50-100\%$.
Jets are taken with $\mathrm {p^{jet}_T} > 120$ GeV, $R = 0.4$ and $| \eta_{\rm jet} | < 1.6$.
}
	\label{fig:pT}
\end{figure*}

\subsection{Jet suppression in A$+$A collisions \label{subsec:A}}

We first check the nuclear modification factor for inclusive jet productions.
The numerical results from the AMPT model are presented in Fig. \ref{fig:RAA}, compared with the ATLAS data \cite{Aaboud:2018twu}.
Here the nuclear modification factors $R_\mathrm{AA}(\mathrm {p^{jet}_T})$ are plotted as a function of jet transverse momentum $\mathrm {p^{jet}_T}$ in $0-10\%$ and $70-80\%$ Pb+Pb collisions at $\sqrt {s_\mathrm{NN}} = 5.02$ TeV.
We use the same kinematic cuts as the ATLAS measurements: jet cone size $R = 0.4$, jet pseudorapidity cut $\left| y \right| < 2.8$.
In the ATLAS data, the error bars represent statistical uncertainties, and the shaded boxes around the data points represent bin-wise correlated systematic uncertainties.

A significant suppression of jet productions in central Pb+Pb collisions relative to p+p collisions is observed.
As jet $p_\mathrm{T}$ increases, the nuclear modification factor $R_\mathrm{AA}(\mathrm {p^{jet}_T})$ increases.
This result indicates jets may experience a significant amount of energy loss due to the interactions with the bulk matter when they propagate through the hot QGP produced in Pb+Pb collisions.
In central collisions with denser and larger media, the effect from jet-medium interaction is larger, leading to larger jet energy loss and thus smaller $R_\mathrm{AA}(\mathrm {p^{jet}_T})$ compared to peripheral collisions.

In the above $R_\mathrm{AA}(\mathrm {p^{jet}_T})$ calculations, we have chosen the same value of partonic cross section $\sigma = 1.5$~mb as in our previous work \cite{Gao:2016ldo}, where a reasonable description of the momentum imbalance and dijet asymmetry data from CMS \cite{CMS:2011iwn} is obtained.
Therefore, the same value $\sigma = 1.5$~mb will be used for studying the correlations between charged particles and reconstructed jets presented in the following subsections.

\begin{figure*}
	\centering
	\includegraphics[width=0.72\linewidth]{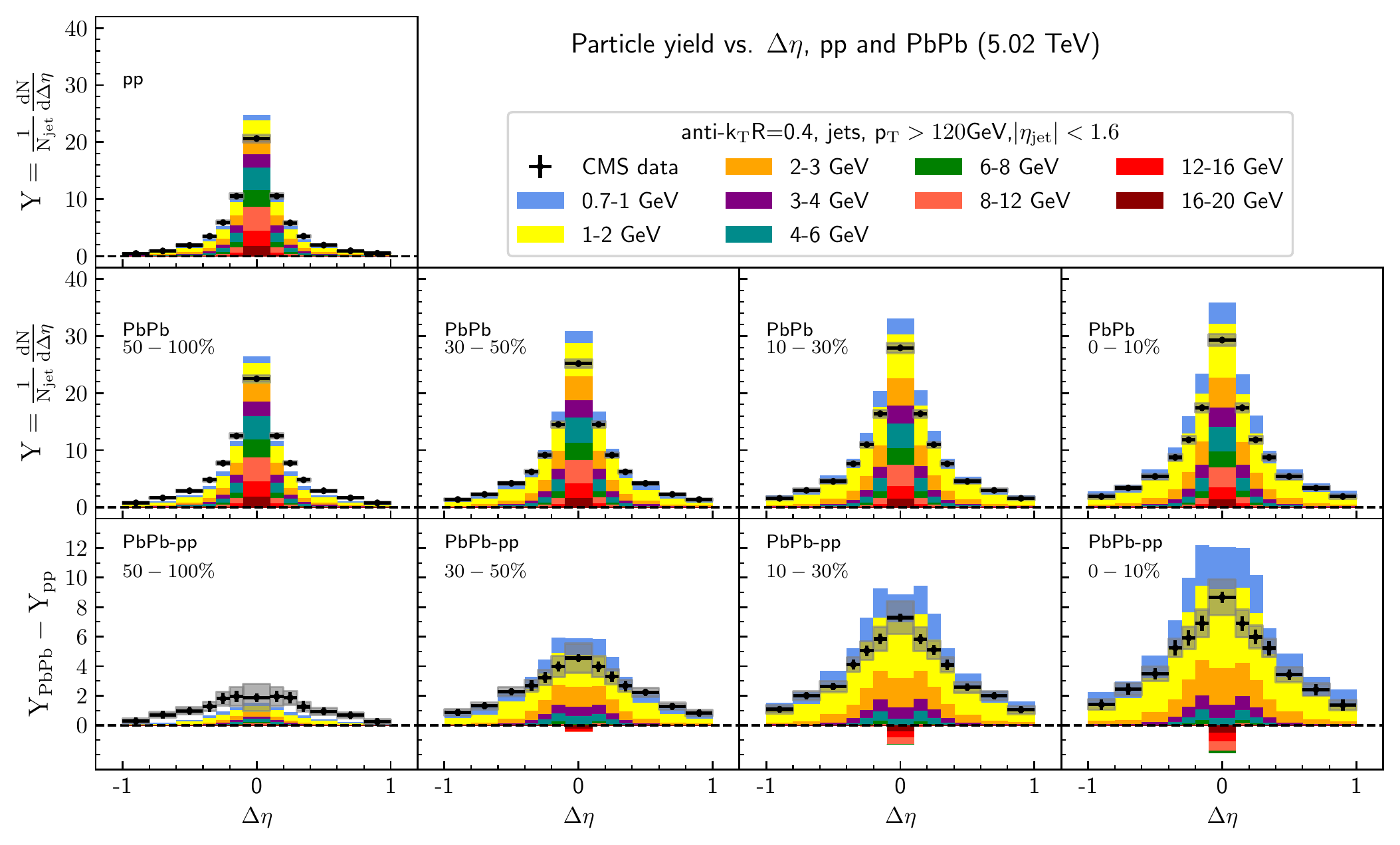}
	\caption{Jet-correlated charged particle yield distribution as a function of $\Delta \eta$ in p+p (top left panel), Pb+Pb (middle panels) collisions and their differences (lower panels).
Four different Pb+Pb collision centralities are shown: $0-10\%, 10-30\%, 30-50\%$, and $50-100\%$.
Jets are taken with $\mathrm {p^{jet}_T} > 120$ GeV, $R = 0.4$ and $| \eta_{\rm jet} | < 1.6$.
The stacked histograms are obtained from the AMPT model: different colors represent the contributions from different $p_\mathrm{T}$ intervals for charged particles.
The accumulated histogram for $0.7 < p_\mathrm{T} < 20$~GeV is compared to CMS data \cite{CMS:2018zze}.}
	\label{fig:eta}
\end{figure*}

\subsection{$p_\mathrm{T}$ distribution of the jet-correlated charged particles}

Let us now investigate the correlations between charged particles and reconstructed jets.
Following CMS Collaboration, we take the reconstructed jets with $\mathrm {p^{jet}_T} > 120$ GeV, $R = 0.4$ and $| \eta_{\rm jet} | < 1.6$ and look at charged particles in the region $\Delta r < 1$ around the jet axis.
Using the three-dimensional charged particle distribution $\frac{\mathrm d ^3 N}{\mathrm dp_\mathrm T \mathrm{d} \Delta \phi \mathrm{d} \Delta \eta}$ around the jets, we may integrate out the pseudorapidity and angular parts ($\Delta \eta$ and $\Delta \phi$) to obtain the $p_{\rm T}$ distributions of charged particles around reconstructed jets:
\begin{equation}
	\frac{\mathrm d N}{\mathrm dp_\mathrm T} = \int \mathrm{d} \Delta \phi \int \mathrm{d} \Delta \eta
	\frac{\mathrm d ^3 N}{\mathrm dp_\mathrm T \mathrm{d} \Delta  \phi \mathrm{d} \Delta \eta } \bigg| _{\Delta r < 1}.
\end{equation}
In Fig. \ref{fig:pT}, we show jet-correlated charged particle $p_{\rm T}$ distributions around jets from the AMPT model and compared to the CMS data. The upper panel shows the results for for Pb+Pb and p+p collisions, and the lower panel show their difference.
One can see that our calculation can provide a reasonable description of the CMS data.
The difference between Pb+Pb and p+p events for low-$p_\mathrm{T}$ particles shows strong dependence on collision centrality.
For more central collisions, more soft particles are observed around the jets (within $\Delta r < 1$).
This is due to larger jet energy loss in more central collisions, which leads to more low $p_{\rm T}$ particles around jets.
In contrast, the yield of high $p_{\rm T}$ particles around jets in Pb+Pb collisions are suppressed compared to p+p collisions.
The above result is consistent with the $\Delta r$ distribution of charged particles around jets (see later).

In the above charged particle $p_{\rm T}$ distribution around jets, the pseudorapidity and angular parts ($\Delta \eta$, $\Delta \phi$) of the distribution have been integrated out.
In the following, we will study in detail how charged particles of different $p_{\rm T}$ contribute to the $\Delta \eta$, $\Delta \phi$ and $\Delta r$ distributions of charged particles with respect to the jet axis.

\begin{figure*}
	\centering
	\includegraphics[width=0.72\linewidth]{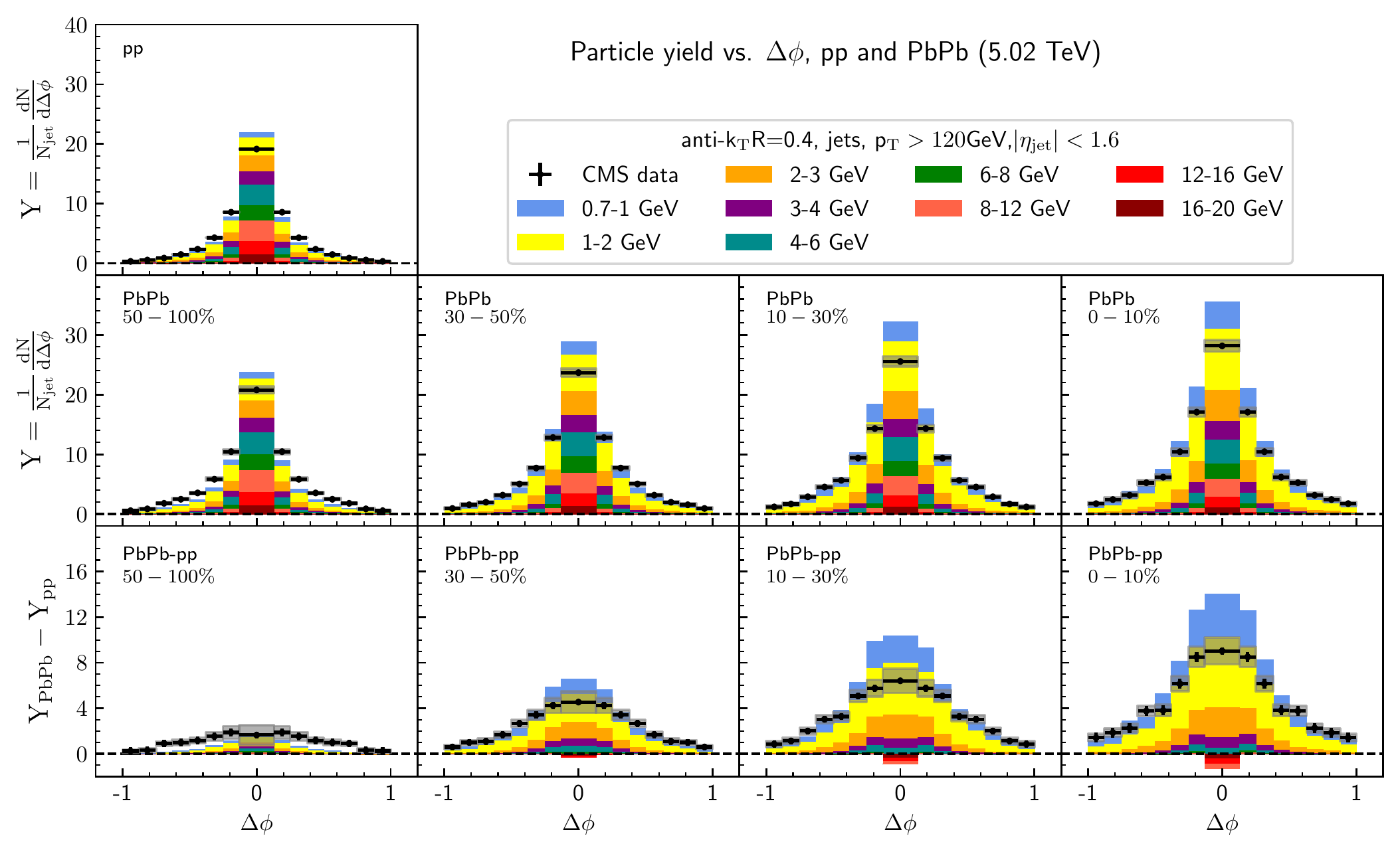}
	\caption{Jet-correlated charged particle yield distribution as a function of $\Delta \phi$ in p+p (top left panel), Pb+Pb (middle panels) collisions and their differences (lower panels).
Four different Pb+Pb collision centralities are shown: $0-10\%, 10-30\%, 30-50\%$, and $50-100\%$.
Jets are taken with $\mathrm {p^{jet}_T} > 120$ GeV, $R = 0.4$ and $| \eta_{\rm jet} | < 1.6$.
The stacked histograms are obtained from the AMPT model: different colors represent the contributions from different $p_\mathrm{T}$ intervals for charged particles.
The accumulated histogram for $0.7 < p_\mathrm{T} < 20$~GeV is compared to CMS data \cite{CMS:2018zze}.}
	\label{fig:phi}
\end{figure*}

\begin{figure*}
	\centering
	\includegraphics[width=0.72\linewidth]{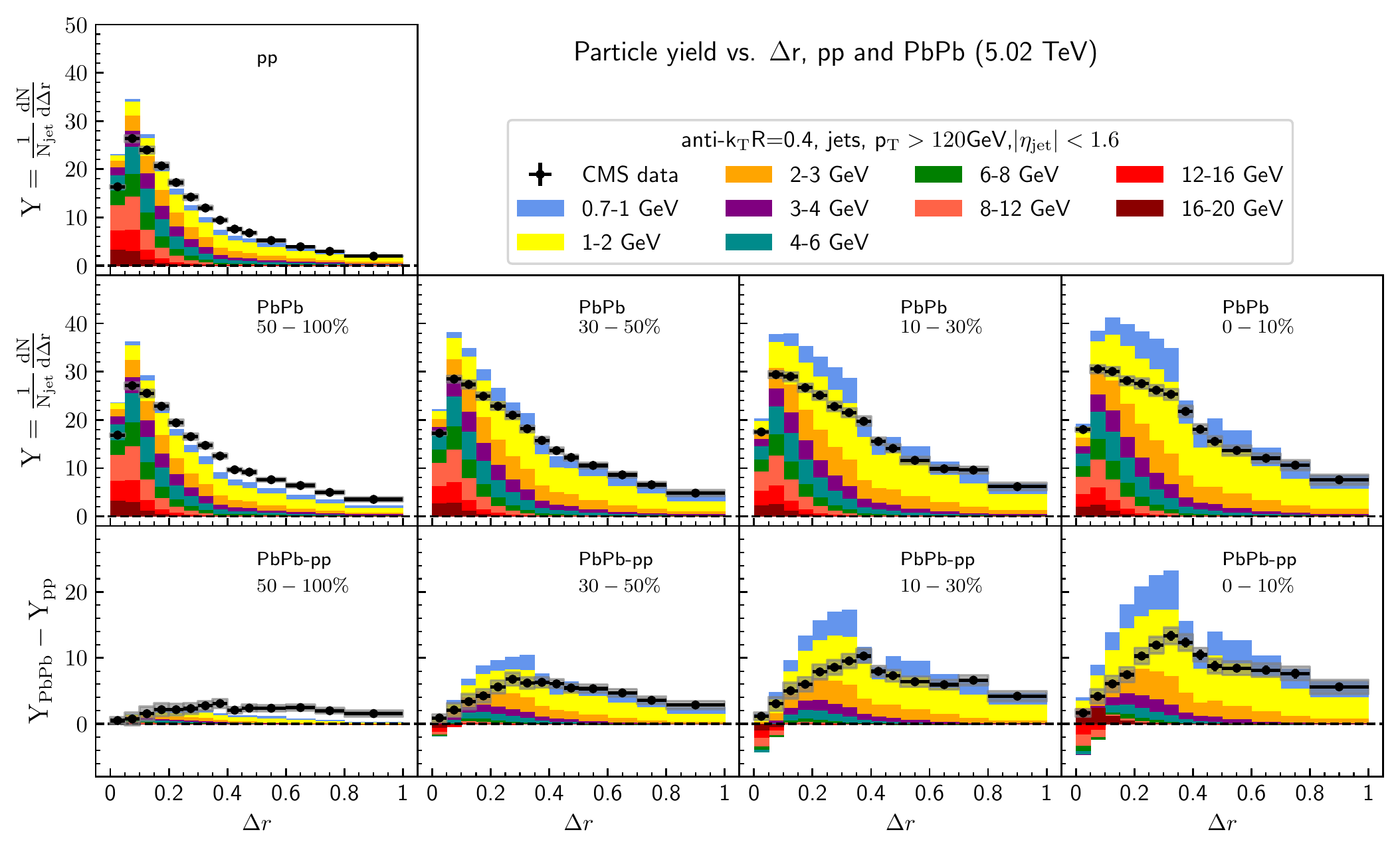}
	\caption{Jet-correlated charged particle yield distribution as a function of $\Delta r$ in p+p (top left panel), Pb+Pb (middle panels) collisions and their differences (lower panels).
Four different Pb+Pb collision centralities are shown: $0-10\%, 10-30\%, 30-50\%$, and $50-100\%$.
Jets are taken with $\mathrm {p^{jet}_T} > 120$ GeV, $R = 0.4$ and $| \eta_{\rm jet} | < 1.6$.
The stacked histograms are obtained from the AMPT model: different colors represent the contributions from different $p_\mathrm{T}$ intervals for charged particles.
The accumulated histogram for $0.7 < p_\mathrm{T} < 20$~GeV is compared to CMS data \cite{CMS:2018zze}.}
	\label{fig:DeltaR}
\end{figure*}

\begin{figure*}
	\centering
	\includegraphics[width=0.72\linewidth]{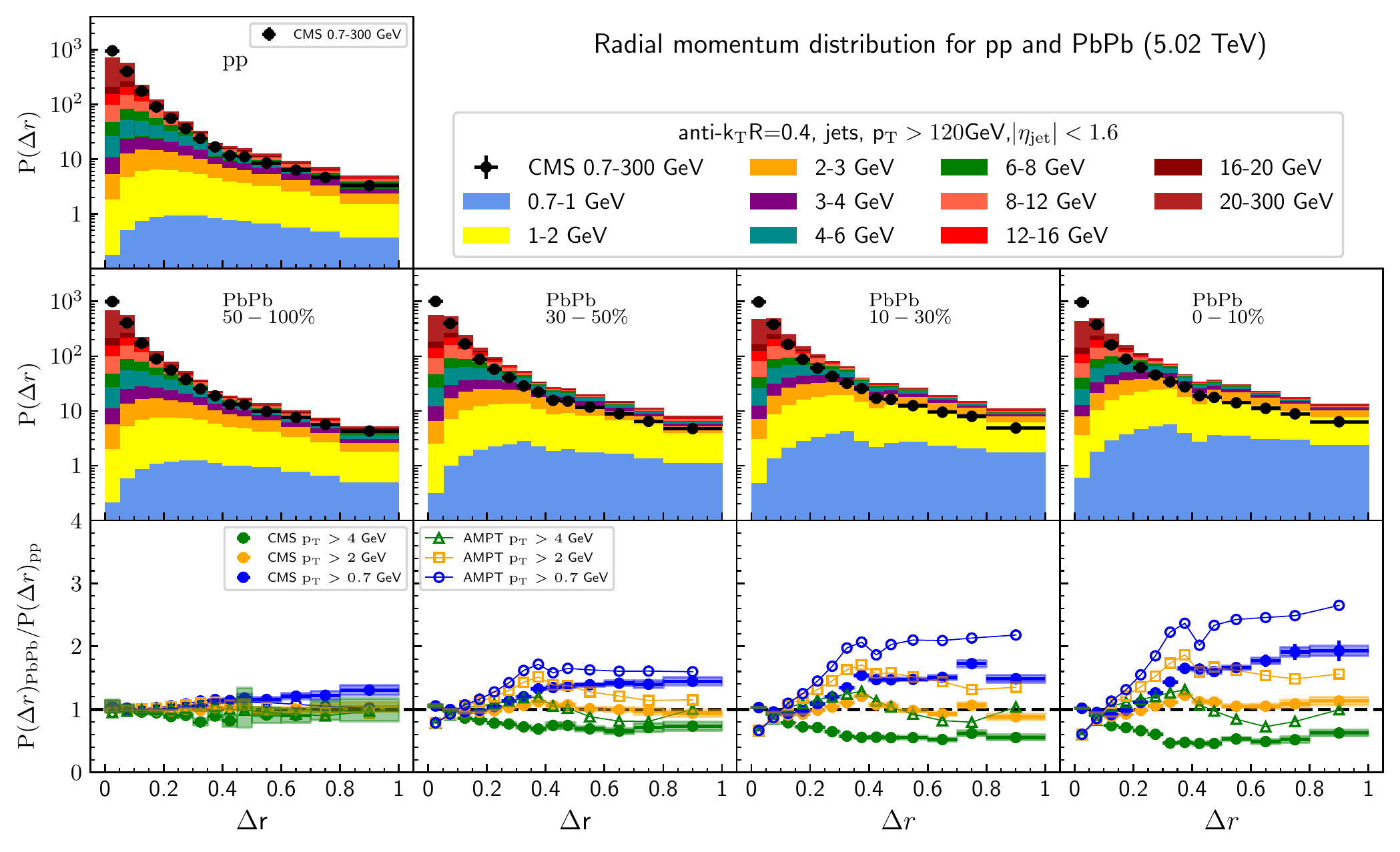}
	\caption{Jet transverse momentum profile $ P(\Delta r)$ for jets with $\mathrm {p^{jet}_T} > 120$ GeV, $R = 0.4$ and $| \eta_{\rm jet} | < 1.6$ in p+p (top left panel) and Pb+Pb (middle panels) collisions.
Four different Pb+Pb collision centralities are shown: $0-10\%, 10-30\%, 30-50\%$, and $50-100\%$.
The stacked histograms are obtained from the AMPT model: different colors represent the contributions from different $p_\mathrm{T}$ intervals for charged particles.
The accumulated histogram is for $0.7 < p_\mathrm{T} < 300$~GeV.
The bottom panels show the ratio between Pb+Pb and p+p collisions for different $p_\mathrm{T}$ intervals, $4-300$~GeV, $2-300$~GeV and $0.7-300$~GeV.
The data are taken from CMS \cite{CMS:2018zze}. }
	\label{fig:P}
\end{figure*}

\begin{figure*}
	\centering
	\includegraphics[width=0.72\linewidth]{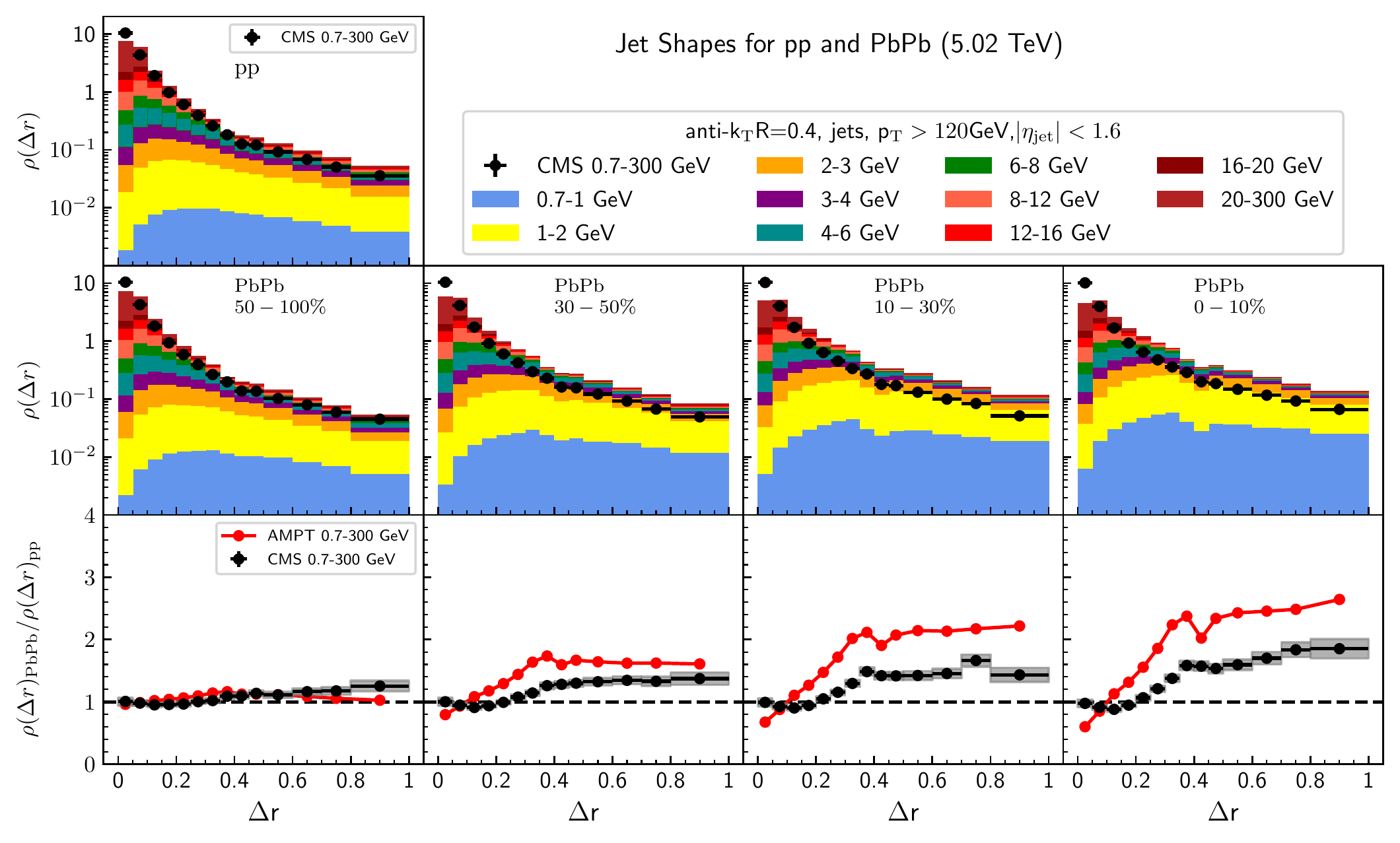}
	\caption{Jet shape $\rho(\Delta r)$ for jets with $\mathrm {p^{jet}_T} > 120$ GeV, $R = 0.4$ and $| \eta_{\rm jet} | < 1.6$ in p+p (top left panel) and Pb+Pb (middle panels) collisions.
Four different Pb+Pb collision centralities are shown: $0-10\%, 10-30\%, 30-50\%$, and $50-100\%$.
The stacked histograms are obtained from the AMPT model: different colors represent the contributions from different $p_\mathrm{T}$ intervals for charged particles.
The accumulated histogram is for $0.7 < p_\mathrm{T} < 300$~GeV.
The bottom panels show the ratio between Pb+Pb and p+p collisions for $p_\mathrm{T}=0.7-300$~GeV.
The data are taken from CMS \cite{CMS:2018zze}.}
	\label{fig:rho}
\end{figure*}

\begin{figure}
	\centering
	\includegraphics[width=1\linewidth]{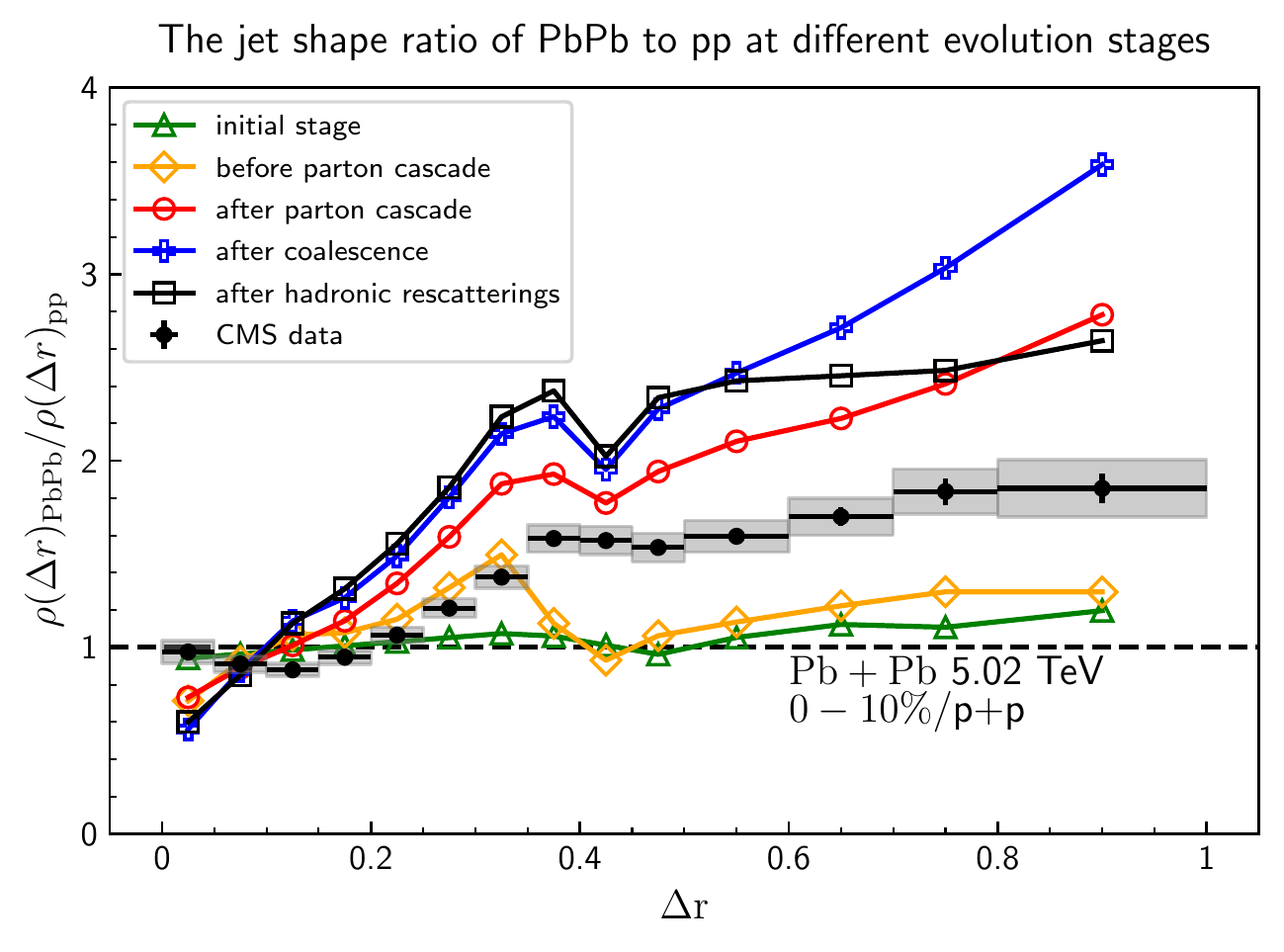}
	\caption{The ratio of jet shape function $\rho(\Delta r)$ in Pb+Pb collisions to that in p+p collisions for different scenarios: initial stage, before parton cascade, after parton cascade, after coalescence, and after hadronic rescatterings. The data are taken from CMS \cite{CMS:2018zze}. }
	\label{fig:evolution_stage}
\end{figure}

\subsection{$\Delta \eta$, $\Delta \phi$ and $\Delta r$ distributions of the jet-correlated charged particles \label{subsection:C}}

Now we look at the $\Delta \eta$, $\Delta \phi$ and $\Delta r$ distributions of charge particles around reconstructed jets.
Given the three-dimensional distribution $\frac{\mathrm d ^3 N}{\mathrm dp_\mathrm T \mathrm{d} \Delta \phi \mathrm{d} \Delta \eta}$, one may first perform the integration over $\Delta \phi$ or $\Delta \eta$ to obtain the following pseudorapidity or azimuthal angle distributions:
\begin{equation}
	\frac{\mathrm d N}{\mathrm d\Delta \eta} = \int \mathrm{d} \Delta \phi \int dp\mathrm{_ T}
	\frac{\mathrm d ^3 N}{\mathrm dp_\mathrm T \mathrm{d} \Delta \phi \mathrm{d} \Delta \eta } \bigg| _{|\Delta \phi| < 1},
\end{equation}
\begin{equation}
	\frac{\mathrm d N}{\mathrm d\Delta \phi} = \int \mathrm{d} \Delta \eta \int dp\mathrm{_ T}
	\frac{\mathrm d ^3 N}{\mathrm dp_\mathrm T \mathrm{d} \Delta \phi \mathrm{d} \Delta \eta } \bigg| _{|\Delta \eta| < 1}.
\end{equation}
In Fig. \ref{fig:eta} and Fig. \ref{fig:phi}, we show the charged particle $\Delta \eta$ and $\Delta \phi$ distributions around reconstructed jets.
The top left panel shows the result for p+p collisions, the middle panels show the results for Pb+Pb collisions, and the lower panels show the differences between Pb+Pb and p+p collisions.
In the analysis, jets are taken with $\mathrm {p^{jet}_T} > 120$ GeV, $R = 0.4$ and $| \eta_{\rm jet} | < 1.6$.
We have studied Pb+Pb collisions for four different centralities: $0-10\%, 10-30\%, 30-50\%$, and $50-100\%$.
In the figure, the stacked histograms are the results obtained from the AMPT model; different colors represent the contributions from charged particles in several different $p_\mathrm{T}$ intervals bounded by 0.7, 1, 2, 3, 4, 6, 8, 12, 16 and 20 GeV.
The accumulated histogram represents the result for $0.7 < p_\mathrm{T} < 20$ GeV, which is compared to CMS data \cite{CMS:2018zze}.

From Figs. \ref{fig:eta} and \ref{fig:phi} we find that our AMPT calculations for both $\Delta \eta$ and $\Delta \phi$ distributions with $0.7-20$~GeV in p+p collisions can qualitatively describe the CMS data.
As for Pb+Pb collisions, the AMPT model can qualitatively describe the main feature of the CMS data.
Focusing on the difference between the Pb+Pb and p+p collisions, we see a depletion of hard particles ($p_\mathrm{T}>6-8$~GeV) and an enhancement of soft particles ($p_\mathrm{T}<4-6$~GeV) in Pb+Pb collisions compared to the p+p collisions.
Such effect is stronger in central Pb+Pb collisions than in peripheral collisions, and is symmetric in both $\Delta \eta$ and $\Delta \phi$.
One interesting feature is that the enhancement of soft particles extends to very large $\Delta\eta$ and $\Delta \phi$, which means that the lost energy from jets is transported to soft particles at very large angles.
The enhancement of soft particles at large $\Delta\eta$ or $\Delta \phi$ from the AMPT model is larger than the CMS data.
This difference might be due to the fact that the AMPT model only includes elastic collisions and the medium-induced radiative processes are missing in the model.
For similar jet energy loss and jet suppression factor, elastic collisions are usually more effective than medium-induced radiation in transporting the energy and momentum to larger angles.
The inclusion of inelastic radiative processes is expected to improve the agreement with the data.

One may further study the $\Delta r$ distribution of charged particles around jets, which can be obtained as follows:
\begin{align}
	\frac{\mathrm d N}{\mathrm d\Delta r} &= \int \mathrm{d} \Delta \phi \int \mathrm{d} \Delta \eta \int dp\mathrm{_ T}
	\frac{\mathrm d ^3 N}{\mathrm dp_\mathrm T \mathrm{d} \Delta \phi \mathrm{d} \Delta \eta }
\nonumber\\
& \times \delta(\Delta r - \sqrt{(\Delta\phi)^2 + (\Delta\eta)^2}).
\end{align}
In practice, we construct the above charged particle distribution in annular rings of width $\delta r = 0.05$ around jet axis.
Fig. \ref{fig:DeltaR} shows jet-correlated charged particle yield distribution as a function of $\Delta r$. The top left panel shows the result for p+p collisions, the middle for Pb+Pb collisions, and the lower for the differences between Pb+Pb and p+p collisions. The labels and setups are similar to Fig. \ref{fig:eta}.
One can see that jet-correlated charged particle yield distributions as a function of $\Delta r$ for $0.7-20$~GeV in p+p collisions from the AMPT calculation agrees with CMS data qualitatively.
Comparing Pb+Pb and p+p collisions, we observe a depletion of hard particles ($p_\mathrm{T}>6-8$~GeV) and an enhancement of soft particles ($p_\mathrm{T}<4-6$~GeV).
Such effect is stronger in central Pb+Pb collisions than in peripheral collisions.
The enhancement of soft particles extends to very large $\Delta r$, since the lost energy from jets can be transported to soft particles at very large angles.
Again, the discrepancy between the AMPT model and CMS data for Pb+Pb collisions is because the AMPT model does not have medium-induced radiative processes.

\subsection{Jet $p_ \mathrm{T}$ profile and jet shape \label{subsection:D}}

In the previous subsection, we have studied the $p_T$, $\Delta \eta$, $\Delta \phi$ and $\Delta r$ distributions of charged particle yield around the reconstructed jets.
Using the three-dimensional distribution $\frac{\mathrm d^3N}{\mathrm dp_{\rm T} \mathrm d\Delta \eta \mathrm d\Delta \phi}$ for charged particles with respect to jet axis, we may construct jet transverse momentum profile $ P(\Delta r)$ as follows:
\begin{align}
	{P}(\Delta r) &= \frac{1}{N_\mathrm{jets}} \int \mathrm{d} \Delta \phi \int \mathrm{d} \Delta \eta \int \mathrm{d}p\mathrm{_ T}
		\frac{\mathrm d ^3 N}{\mathrm dp_\mathrm T \mathrm{d} \Delta \phi \mathrm{d} \Delta \eta } p_{\rm T}
\nonumber\\
& \times \delta(\Delta r - \sqrt{(\Delta\phi)^2 + (\Delta\eta)^2}).
\end{align}
In practice, $ P(\Delta r)$ is just the distribution of charged particles in the annular rings around the jet axis with each particle weighted by its transverse momentum, see Eq. (\ref{equation:P}).
In Fig. \ref{fig:P}, we show jet transverse momentum profile $ P(\Delta r)$ in p+p collisions (top left panel) and Pb+Pb collisions (middle panels).
The Pb+Pb results are shown for four centralities: $0-10\%, 10-30\%, 30-50\%$, and $50-100\%$.
The stacked histograms are results from the AMPT model: different colors represent the contributions from different $p_\mathrm{T}$ intervals for charged particles.
The accumulated histogram is for $0.7 < p_\mathrm{T} < 20$~GeV.
The bottom panels show the ratio between Pb+Pb and p+p collisions for different $p_\mathrm{T}$ intervals, $4-300$~GeV, $2-300$~GeV and $0.7-300$~GeV.

From Fig. \ref{fig:P}, one observes a large excess of soft particles together with a depletion of hard particles in Pb+Pb collisions relative to p+p collisions.
The enhancement of soft particles is more prominent at large $\Delta r$ region.
This means that the energy is transported from hard jets to soft particles at large $\Delta r$ regions away from jet axis.
Such effect is stronger in central collisions than in peripheral collisions.
We also see that our AMPT model calculations overestimate the accumulated transverse momentum profile at large $\Delta r$ region, which may be due to the negligence of inelastic radiative process in the parton cascade.

One may normalize the radial distribution $P(\Delta r)$ of jet transverse momentum profile and further investigate the jet shape function $\rho(\Delta r)$.
Fig. \ref{fig:rho} shows jet shape $\mathrm \rho (\Delta r)$ for jets with $\mathrm {p^{jet}_T} > 120$ GeV, $R = 0.4$ and $| \eta_{\rm jet} | < 1.6$ in p+p (top left panel) and Pb+Pb (middle panels) collisions.
The stacked histograms are obtained from the AMPT model: different colors represent the contributions from different $p_\mathrm{T}$ intervals for charged particles.
The bottom panels show the ratio between Pb+Pb and p+p collisions for $p_\mathrm{T}=0.7-300$~GeV.

Our numerical results for jet shape $\rho(\Delta r)$ show a clear centrality dependence in Pb+Pb collisions.
In most peripheral Pb+Pb collisions, jet shape $\rho (\Delta r)$ is similar to that in p+p collisions.
In central Pb+Pb collisions, the ratio of jet shape function in Pb+Pb collisions and p+p collisions show a strong enhancement at large values of $\Delta r$ together with a depletion at smaller values of $\Delta r$.
This indicates that a significant fraction of the energy (transverse momentum) is transported to larger angles with respect to the jet axis.
For very small values of $\Delta r$, jet shape function in Pb+Pb collisions from the AMPT model shows some suppression effect as compared to p+p collisions, while the CMS data show much smaller nuclear modification effect.
This difference might be due to the lack of medium-induced radiation in jet-medium interaction in the AMPT model.

To see how different stages of the collisions contribute to the modification of jet shape, in Fig. \ref{fig:evolution_stage} we show the ratio of jet shape in Pb+Pb collision and p+p at different evolution stages in the AMPT model.
One can see that before parton cascade, there is some change in jet shape function as compared to p+p collisions, which is due to the initial scatterings in HIJING.
Parton cascade stage in the AMPT model gives a very significant contribution to the change of jet shape function. 
Due to the elastic scatterings between jets and medium partons, the energy and momentum are transported to large distance away from jets.
Jet shape function gets mild change in the coalescence and hadronic rescattering processes.

\section{Summary \label{sec:summary}}

Within the framework of a multi-phase transport model, we have studied jet suppression and the redistribution of jet energy loss in relativistic heavy-ion collisions.
Following the CMS Collaboration, we have performed a detailed study on the correlations between charged particles and reconstructed jets in p+p and Pb+Pb collisions at $\sqrt {s_\mathrm{NN}} = 5.02$ TeV.
The charged particle yield distribution with respect to jet axis as a function of $\Delta \eta$, $\Delta \phi$ are studied for different collision centralities and for different transverse momenta of charged particles.
It is found that the charged particle distributions around the reconstructed jets exhibit a strong enhancement of soft particles in Pb+Pb collisions relative to p+p collisions.
Such enhancement extends from intermediate to very large values of $\Delta r$.
By weighting each particle with its $p_ \mathrm T$ value, we also study jet transverse momentum profile and jet shape function using the correlations between charged particles with respect to jet axis.
A redistribution of energy from small to large angles with respect to jet axis is observed in Pb+Pb collisions compared to p+p collisions at the LHC.
The above effects are more prominent for more central Pb+Pb collisions.
Our study indicates that the lost energy from the hard jets is transported from small to large angles away from jet axis and carried by soft particles in the final state of the nuclear collisions.
The numerical results from the AMPT model are qualitatively consistent with the CMS data.
More quantitatively, the AMPT model overestimates the enhancement of soft particles at large $\Delta r$ compared to the CMS data.
This may be due to the fact that the AMPT model only includes elastic collisions between jet partons and medium partons in the parton cascade stage.
The inclusion of inelastic radiative processes in the parton cascade is expected to improve the agreement with the experimental data.
The study along this direction will be left for future effort.

\section{ACKNOWLEDGMENTS}

We thank X.-N. Wang for helpful discussions. This work is supported by National Natural Science Foundation of China under Grants No. 11935007,  No. 11775095, No. 11890710, No. 11890711, and in part by the Guangdong Major Project of Basic and Applied Basic Research (No. 2020B0301030008).

\bibliographystyle{h-physrev5}
\bibliography{refs_GYQ}% Produces the bibliography via BibTeX.

\begin{thebibliography}{10}

\bibitem{Gyulassy:2004zy}
M.~Gyulassy and L.~McLerran,
\newblock Nucl. Phys. A {\bf 750}, 30 (2005), arXiv:nucl-th/0405013.

\bibitem{Adcox:2004mh}
PHENIX, K.~Adcox {\em et~al.},
\newblock Nucl. Phys. {\bf A757}, 184 (2005), arXiv:nucl-ex/0410003.
%%CITATION = NUCL-EX/0410003;%%

\bibitem{Arsene:2004fa}
BRAHMS, I.~Arsene {\em et~al.},
\newblock Nucl. Phys. {\bf A757}, 1 (2005), arXiv:nucl-ex/0410020.
%%CITATION = NUCL-EX/0410020;%%

\bibitem{Back:2004je}
B.~B. Back {\em et~al.},
\newblock Nucl. Phys. {\bf A757}, 28 (2005), arXiv:nucl-ex/0410022.
%%CITATION = NUCL-EX/0410022;%%

\bibitem{Adams:2005dq}
STAR, J.~Adams {\em et~al.},
\newblock Nucl. Phys. {\bf A757}, 102 (2005), arXiv:nucl-ex/0501009.
%%CITATION = NUCL-EX/0501009;%%

\bibitem{Muller:2006ee}
B.~Muller and J.~L. Nagle,
\newblock Ann. Rev. Nucl. Part. Sci. {\bf 56}, 93 (2006),
  arXiv:nucl-th/0602029.
%%CITATION = NUCL-TH/0602029;%%

\bibitem{Muller:2012zq}
B.~Muller, J.~Schukraft, and B.~Wyslouch,
\newblock Ann. Rev. Nucl. Part. Sci. {\bf 62}, 361 (2012), arXiv:1202.3233.

\bibitem{Wang:1991xy}
X.-N. Wang and M.~Gyulassy,
\newblock Phys.Rev.Lett. {\bf 68}, 1480 (1992).
%%CITATION = PRLTA,68,1480;%%

\bibitem{Qin:2015srf}
G.-Y. Qin and X.-N. Wang,
\newblock Int. J. Mod. Phys. E {\bf 24}, 1530014 (2015), arXiv:1511.00790.

\bibitem{Cao:2020wlm}
S.~Cao and X.-N. Wang,
\newblock Rept. Prog. Phys. {\bf 84}, 024301 (2021), arXiv:2002.04028.

\bibitem{Bernhard:2019bmu}
J.~E. Bernhard, J.~S. Moreland, and S.~A. Bass,
\newblock Nature Phys. {\bf 15}, 1113 (2019).

\bibitem{Everett:2020xug}
JETSCAPE, D.~Everett {\em et~al.},
\newblock Phys. Rev. C {\bf 103}, 054904 (2021), arXiv:2011.01430.

\bibitem{Burke:2013yra}
JET, K.~M. Burke {\em et~al.},
\newblock Phys. Rev. {\bf C90}, 014909 (2014), arXiv:1312.5003.
%%CITATION = ARXIV:1312.5003;%%

\bibitem{Cao:2021keo}
S.~Cao {\em et~al.},
\newblock (2021), arXiv:2102.11337.

\bibitem{Adcox:2001jp}
PHENIX, K.~Adcox {\em et~al.},
\newblock Phys. Rev. Lett. {\bf 88}, 022301 (2002), arXiv:nucl-ex/0109003.
%%CITATION = NUCL-EX/0109003;%%

\bibitem{Adler:2002xw}
STAR, C.~Adler {\em et~al.},
\newblock Phys. Rev. Lett. {\bf 89}, 202301 (2002), arXiv:nucl-ex/0206011.
%%CITATION = NUCL-EX/0206011;%%

\bibitem{Aamodt:2010jd}
ALICE Collaboration, K.~Aamodt {\em et~al.},
\newblock Phys.Lett. {\bf B696}, 30 (2011), arXiv:1012.1004.
%%CITATION = ARXIV:1012.1004;%%

\bibitem{Qin:2007rn}
G.-Y. Qin {\em et~al.},
\newblock Phys. Rev. Lett. {\bf 100}, 072301 (2008), arXiv:0710.0605.
%%CITATION = 0710.0605;%%

\bibitem{Zhang:2007ja}
H.~Zhang, J.~F. Owens, E.~Wang, and X.-N. Wang,
\newblock Phys. Rev. Lett. {\bf 98}, 212301 (2007), arXiv:nucl-th/0701045.
%%CITATION = NUCL-TH/0701045;%%

\bibitem{Zhang:2009rn}
H.~Zhang, J.~F. Owens, E.~Wang, and X.-N. Wang,
\newblock Phys. Rev. Lett. {\bf 103}, 032302 (2009), arXiv:0902.4000.
%%CITATION = ARXIV:0902.4000;%%

\bibitem{Bass:2008rv}
S.~A. Bass {\em et~al.},
\newblock Phys. Rev. {\bf C79}, 024901 (2009), arXiv:0808.0908.
%%CITATION = 0808.0908;%%

\bibitem{Qin:2009bk}
G.-Y. Qin, J.~Ruppert, C.~Gale, S.~Jeon, and G.~D. Moore,
\newblock Phys.Rev. {\bf C80}, 054909 (2009), arXiv:0906.3280.
%%CITATION = ARXIV:0906.3280;%%

\bibitem{Chen:2011vt}
X.-F. Chen, T.~Hirano, E.~Wang, X.-N. Wang, and H.~Zhang,
\newblock Phys.Rev. {\bf C84}, 034902 (2011), arXiv:1102.5614.
%%CITATION = ARXIV:1102.5614;%%

\bibitem{Chen:2016vem}
L.~Chen, G.-Y. Qin, S.-Y. Wei, B.-W. Xiao, and H.-Z. Zhang,
\newblock Phys. Lett. {\bf B773}, 672 (2017), arXiv:1607.01932.
%%CITATION = ARXIV:1607.01932;%%

\bibitem{Chen:2016cof}
L.~Chen, G.-Y. Qin, S.-Y. Wei, B.-W. Xiao, and H.-Z. Zhang,
\newblock Phys. Lett. {\bf B782}, 773 (2018), arXiv:1612.04202.
%%CITATION = ARXIV:1612.04202;%%

\bibitem{Zhang:2018urd}
S.-L. Zhang, T.~Luo, X.-N. Wang, and B.-W. Zhang,
\newblock Phys. Rev. {\bf C98}, 021901 (2018), arXiv:1804.11041.
%%CITATION = ARXIV:1804.11041;%%

\bibitem{CasalderreySolana:2004qm}
J.~Casalderrey-Solana, E.~V. Shuryak, and D.~Teaney,
\newblock J. Phys. Conf. Ser. {\bf 27}, 22 (2005), arXiv:hep-ph/0411315.

\bibitem{Chaudhuri:2005vc}
A.~K. Chaudhuri and U.~Heinz,
\newblock Phys. Rev. Lett. {\bf 97}, 062301 (2006), arXiv:nucl-th/0503028.

\bibitem{Ruppert:2005uz}
J.~Ruppert and B.~Muller,
\newblock Phys. Lett. B {\bf 618}, 123 (2005), arXiv:hep-ph/0503158.

\bibitem{Qin:2009uh}
G.~Y. Qin, A.~Majumder, H.~Song, and U.~Heinz,
\newblock Phys. Rev. Lett. {\bf 103}, 152303 (2009), arXiv:0903.2255.

\bibitem{Neufeld:2009ep}
R.~B. Neufeld and B.~Muller,
\newblock Phys. Rev. Lett. {\bf 103}, 042301 (2009), arXiv:0902.2950.

\bibitem{Andrade:2014swa}
R.~P.~G. Andrade, J.~Noronha, and G.~S. Denicol,
\newblock Phys. Rev. {\bf C90}, 024914 (2014), arXiv:1403.1789.
%%CITATION = ARXIV:1403.1789;%%

\bibitem{Schulc:2014jma}
M.~Schulc and B.~Tomavsik,
\newblock Phys. Rev. C {\bf 90}, 064910 (2014), arXiv:1409.6116.

\bibitem{Tachibana:2017syd}
Y.~Tachibana, N.-B. Chang, and G.-Y. Qin,
\newblock Phys. Rev. C {\bf 95}, 044909 (2017), arXiv:1701.07951.

\bibitem{Chang:2019sae}
N.-B. Chang, Y.~Tachibana, and G.-Y. Qin,
\newblock Phys. Lett. B {\bf 801}, 135181 (2020), arXiv:1906.09562.

\bibitem{Chen:2017zte}
W.~Chen, S.~Cao, T.~Luo, L.-G. Pang, and X.-N. Wang,
\newblock Phys. Lett. B {\bf 777}, 86 (2018), arXiv:1704.03648.

\bibitem{Chen:2020tbl}
W.~Chen, S.~Cao, T.~Luo, L.-G. Pang, and X.-N. Wang,
\newblock Phys. Lett. B {\bf 810}, 135783 (2020), arXiv:2005.09678.

\bibitem{Yang:2021iib}
Z.~Yang {\em et~al.},
\newblock (2021), arXiv:2101.05422.

\bibitem{Luo:2018pto}
T.~Luo, S.~Cao, Y.~He, and X.-N. Wang,
\newblock (2018), arXiv:1803.06785.
%%CITATION = ARXIV:1803.06785;%%

\bibitem{Park:2018acg}
C.~Park, S.~Jeon, and C.~Gale,
\newblock Nucl. Phys. A {\bf 982}, 643 (2019), arXiv:1807.06550.

\bibitem{KunnawalkamElayavalli:2017hxo}
R.~Kunnawalkam~Elayavalli and K.~C. Zapp,
\newblock JHEP {\bf 07}, 141 (2017), arXiv:1707.01539.
%%CITATION = ARXIV:1707.01539;%%

\bibitem{Brewer:2017fqy}
J.~Brewer, K.~Rajagopal, A.~Sadofyev, and W.~Van Der~Schee,
\newblock JHEP {\bf 02}, 015 (2018), arXiv:1710.03237.
%%CITATION = ARXIV:1710.03237;%%

\bibitem{Neufeld:2008dx}
R.~Neufeld,
\newblock Phys.Rev. {\bf C79}, 054909 (2009), arXiv:0807.2996.
%%CITATION = ARXIV:0807.2996;%%

\bibitem{Bouras:2014rea}
I.~Bouras, B.~Betz, Z.~Xu, and C.~Greiner,
\newblock Phys.Rev. {\bf C90}, 024904 (2014), arXiv:1401.3019.
%%CITATION = ARXIV:1401.3019;%%

\bibitem{Renk:2005si}
T.~Renk and J.~Ruppert,
\newblock Phys.Rev. {\bf C73}, 011901 (2006), arXiv:hep-ph/0509036.
%%CITATION = HEP-PH/0509036;%%

\bibitem{Ma:2010dv}
G.-L. Ma and X.-N. Wang,
\newblock Phys. Rev. Lett. {\bf 106}, 162301 (2011), arXiv:1011.5249.

\bibitem{Qin:2010mn}
G.-Y. Qin and B.~Muller,
\newblock Phys. Rev. Lett. {\bf 106}, 162302 (2011), arXiv:1012.5280,
\newblock [Erratum: Phys. Rev. Lett.108,189904(2012)].
%%CITATION = ARXIV:1012.5280;%%

\bibitem{Chang:2016gjp}
N.-B. Chang and G.-Y. Qin,
\newblock Phys. Rev. {\bf C94}, 024902 (2016), arXiv:1603.01920.
%%CITATION = ARXIV:1603.01920;%%

\bibitem{Chatrchyan:2013kwa}
CMS, S.~Chatrchyan {\em et~al.},
\newblock Phys. Lett. B {\bf 730}, 243 (2014), arXiv:1310.0878.

\bibitem{Chatrchyan:2014ava}
CMS, S.~Chatrchyan {\em et~al.},
\newblock Phys. Rev. C {\bf 90}, 024908 (2014), arXiv:1406.0932.

\bibitem{Aad:2014wha}
ATLAS, G.~Aad {\em et~al.},
\newblock Phys. Lett. B {\bf 739}, 320 (2014), arXiv:1406.2979.

\bibitem{Khachatryan:2015lha}
CMS, V.~Khachatryan {\em et~al.},
\newblock JHEP {\bf 01}, 006 (2016), arXiv:1509.09029.

\bibitem{Young:2011qx}
C.~Young, B.~Schenke, S.~Jeon, and C.~Gale,
\newblock Phys.Rev. {\bf C84}, 024907 (2011), arXiv:1103.5769.
%%CITATION = ARXIV:1103.5769;%%

\bibitem{Dai:2012am}
W.~Dai, I.~Vitev, and B.-W. Zhang,
\newblock Phys. Rev. Lett. {\bf 110}, 142001 (2013), arXiv:1207.5177.
%%CITATION = ARXIV:1207.5177;%%

\bibitem{Wang:2013cia}
X.-N. Wang and Y.~Zhu,
\newblock Phys. Rev. Lett. {\bf 111}, 062301 (2013), arXiv:1302.5874.
%%CITATION = ARXIV:1302.5874;%%

\bibitem{Blaizot:2013hx}
J.-P. Blaizot, E.~Iancu, and Y.~Mehtar-Tani,
\newblock Phys.Rev.Lett. {\bf 111}, 052001 (2013), arXiv:1301.6102.
%%CITATION = ARXIV:1301.6102;%%

\bibitem{Mehtar-Tani:2014yea}
Y.~Mehtar-Tani and K.~Tywoniuk,
\newblock Phys. Lett. {\bf B744}, 284 (2015), arXiv:1401.8293.
%%CITATION = ARXIV:1401.8293;%%

\bibitem{Cao:2017qpx}
S.~Cao and A.~Majumder,
\newblock (2017), arXiv:1712.10055.
%%CITATION = ARXIV:1712.10055;%%

\bibitem{Kang:2017frl}
Z.-B. Kang, F.~Ringer, and I.~Vitev,
\newblock Phys. Lett. {\bf B769}, 242 (2017), arXiv:1701.05839.
%%CITATION = ARXIV:1701.05839;%%

\bibitem{He:2018xjv}
Y.~He {\em et~al.},
\newblock (2018), arXiv:1809.02525.
%%CITATION = ARXIV:1809.02525;%%

\bibitem{Casalderrey-Solana:2016jvj}
J.~Casalderrey-Solana, D.~Gulhan, G.~Milhano, D.~Pablos, and K.~Rajagopal,
\newblock JHEP {\bf 03}, 135 (2017), arXiv:1609.05842.
%%CITATION = ARXIV:1609.05842;%%

\bibitem{Chien:2016led}
Y.-T. Chien and I.~Vitev,
\newblock Phys. Rev. Lett. {\bf 119}, 112301 (2017), arXiv:1608.07283.
%%CITATION = ARXIV:1608.07283;%%

\bibitem{Milhano:2017nzm}
G.~Milhano, U.~A. Wiedemann, and K.~C. Zapp,
\newblock Phys. Lett. B {\bf 779}, 409 (2018), arXiv:1707.04142.

\bibitem{Gao:2016ldo}
Z.~Gao, A.~Luo, G.-L. Ma, G.-Y. Qin, and H.-Z. Zhang,
\newblock Phys. Rev. C {\bf 97}, 044903 (2018), arXiv:1612.02548.

\bibitem{CMS:2011iwn}
CMS, S.~Chatrchyan {\em et~al.},
\newblock Phys. Rev. C {\bf 84}, 024906 (2011), arXiv:1102.1957.

\bibitem{CMS:2018zze}
CMS, A.~M. Sirunyan {\em et~al.},
\newblock JHEP {\bf 05}, 006 (2018), arXiv:1803.00042.

\bibitem{Lin:2004en}
Z.-W. Lin, C.~M. Ko, B.-A. Li, B.~Zhang, and S.~Pal,
\newblock Phys.Rev. {\bf C72}, 064901 (2005), arXiv:nucl-th/0411110.
%%CITATION = NUCL-TH/0411110;%%

\bibitem{Zhang:2005ni}
B.~Zhang, L.-W. Chen, and C.-M. Ko,
\newblock Phys. Rev. C {\bf 72}, 024906 (2005), arXiv:nucl-th/0502056.

\bibitem{Ma:2011uma}
G.-L. Ma and B.~Zhang,
\newblock Phys. Lett. B {\bf 700}, 39 (2011), arXiv:1101.1701.

\bibitem{Wang:1991hta}
X.-N. Wang and M.~Gyulassy,
\newblock Phys. Rev. D {\bf 44}, 3501 (1991).

\bibitem{Gyulassy:1994ew}
M.~Gyulassy and X.-N. Wang,
\newblock Comput. Phys. Commun. {\bf 83}, 307 (1994), arXiv:nucl-th/9502021.

\bibitem{Sjostrand:1993yb}
T.~Sjostrand,
\newblock Comput. Phys. Commun. {\bf 82}, 74 (1994).

\bibitem{Zhang:1997ej}
B.~Zhang,
\newblock Comput. Phys. Commun. {\bf 109}, 193 (1998), arXiv:nucl-th/9709009.

\bibitem{Lin:2001zk}
Z.-w. Lin and C.~M. Ko,
\newblock Phys. Rev. C {\bf 65}, 034904 (2002), arXiv:nucl-th/0108039.

\bibitem{Li:1995pra}
B.-A. Li and C.~M. Ko,
\newblock Phys. Rev. C {\bf 52}, 2037 (1995), arXiv:nucl-th/9505016.

\bibitem{Ma:2013bia}
G.-L. Ma,
\newblock Phys. Lett. B {\bf 724}, 278 (2013), arXiv:1302.5873.

\bibitem{Ma:2013pha}
G.-L. Ma,
\newblock Phys. Rev. C {\bf 87}, 064901 (2013), arXiv:1304.2841.

\bibitem{Ma:2013gga}
G.-L. Ma,
\newblock Phys. Rev. C {\bf 88}, 021902 (2013), arXiv:1306.1306.

\bibitem{Ma:2013yoa}
G.-L. Ma,
\newblock Phys. Rev. C {\bf 89}, 064909 (2014), arXiv:1310.3701.

\bibitem{Ma:2013uqa}
G.-L. Ma,
\newblock Phys. Rev. C {\bf 89}, 024902 (2014), arXiv:1309.5555.

\bibitem{Cacciari:2011ma}
M.~Cacciari, G.~P. Salam, and G.~Soyez,
\newblock Eur. Phys. J. C {\bf 72}, 1896 (2012), arXiv:1111.6097.

\bibitem{Kodolova:2007hd}
O.~Kodolova, I.~Vardanyan, A.~Nikitenko, and A.~Oulianov,
\newblock Eur. Phys. J. C {\bf 50}, 117 (2007).

\bibitem{ATLAS:2018gwx}
ATLAS, M.~Aaboud {\em et~al.},
\newblock Phys. Lett. B {\bf 790}, 108 (2019), arXiv:1805.05635.

\bibitem{Khachatryan:2016erx}
CMS, V.~Khachatryan {\em et~al.},
\newblock JHEP {\bf 02}, 156 (2016), arXiv:1601.00079.

\bibitem{Khachatryan:2016tfj}
CMS, V.~Khachatryan {\em et~al.},
\newblock JHEP {\bf 11}, 055 (2016), arXiv:1609.02466.

\bibitem{Aaboud:2018twu}
ATLAS, M.~Aaboud {\em et~al.},
\newblock Phys. Lett. B {\bf 790}, 108 (2019), arXiv:1805.05635.

\end{thebibliography}

\end{document}